\documentclass{PoS}
\usepackage{siunitx}
\usepackage{lineno}

\title{Characterization of Two PMT Models for the IceCube Upgrade mDOM}

\ShortTitle{IceCube Upgrade mDOM PMTs}

\author{
The IceCube Collaboration\footnote{For collaboration list, see PoS(ICRC2019) 1177.}\\
{\itshape \href{http://icecube.wisc.edu/collaboration/authors/icrc19_icecube}{http://icecube.wisc.edu/collaboration/authors/icrc19\_icecube}}\\

E-mail: \email{daan.vaneijk@icecube.wisc.edu,judith.js.schneider@fau.de,\\
m.unland@uni-muenster.de}
}

\abstract{

The IceCube Upgrade  will expand the IceCube Neutrino Observatory with nearly 800 new optical modules. A large fraction of these will be multi-PMT optical modules (mDOMs), featuring 24 PMTs pointing uniformly in all directions, providing an almost homogeneous angular coverage and providing an effective photosensitive area more than twice that of current IceCube optical modules. Two PMT models from different manufacturers are currently considered for use in the mDOM: a 3.5 inch PMT from HZC Photonics and a 3 inch PMT from Hamamatsu. Both PMTs have been characterized in terms of gain, timing, quantum efficiency and dark noise rate as a function of temperature. The obtained characterization results are presented here.\\

\vspace{4mm}
{\bfseries Corresponding authors:}
D.\,van Eijk$^{1}$, J.\,Schneider$^{2}$, M.\,Unland$^{3}$\\
{$^{1}$ \itshape WIPAC, UW Madison}\\
{$^{2}$ \itshape ECAP, Friedrich-Alexander-Universit\"at Erlangen-N\"urnberg}\\
{$^{3}$ \itshape Institut f\"ur Kernphysik, Westf\"alische Wilhelms-Universit\"at M\"unster}

}

\FullConference{36th International Cosmic Ray Conference -ICRC2019-\\
		July 24th - August 1st, 2019\\
		Madison, WI, U.S.A.}

\begin{document}

\section{Introduction}\label{sec:introduction}
The IceCube Neutrino Observatory is a cubic-kilometer neutrino detector installed in the ice at the geographic South Pole~\cite{IceCube} between depths of 1450 m and 2450 m. Construction of the full detector was completed in 2010. Reconstruction of the direction, energy and flavor of the neutrinos relies on the optical detection of Cherenkov radiation emitted by charged particles produced via neutrino interactions in the surrounding ice or the nearby bedrock.

\begin{figure}[!h]
\hfill
\includegraphics[trim=0 0 0 2cm, width=0.25\textwidth,clip]{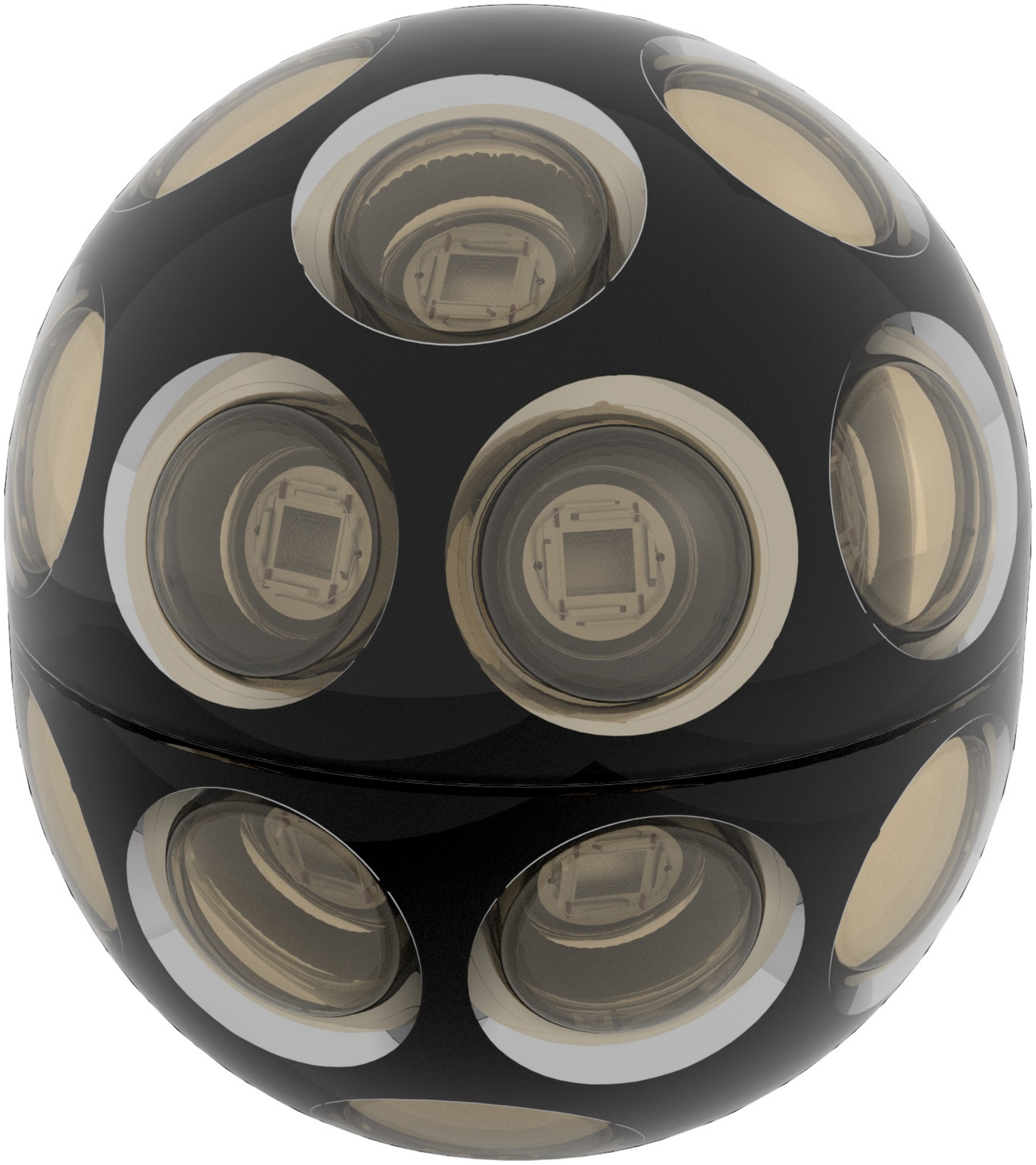}
\hfill
\includegraphics[trim = 7cm 10cm 7cm 10cm, width=0.35\textwidth,clip]{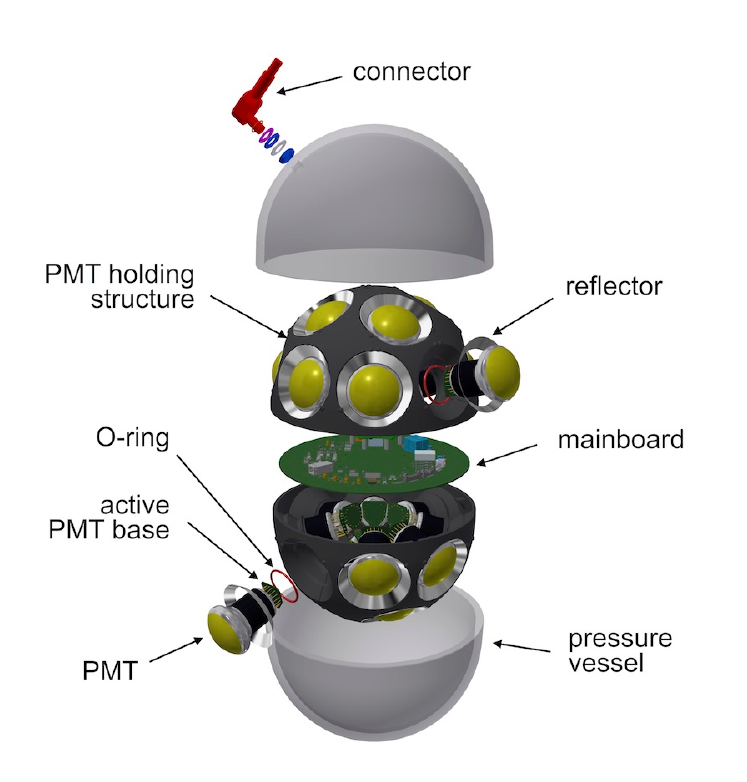}
\hfill
\hfill
\caption{The IceCube Upgrade mDOM design. \emph{Left}: Rendered view. \emph{Right}: Exploded view.}
\label{fig:mDOM}  
\end{figure}

The IceCube Upgrade \cite{IceCubeUpgrade} will consist of nearly 800 new optical modules integrated on 7 vertical strings. A large fraction of the optical modules contain multiple photomultiplier tubes (PMT). One particular design for IceCube Upgrade optical modules, shown in figure \ref{fig:mDOM}, is inspired by the KM3NeT optical module \cite{KM3NeT, KM3NeTDOM} and is called the mDOM. This multi-PMT optical module features 24 PMTs pointing uniformly in all directions, providing an almost homogeneous angular coverage and providing an effective photosensitive area more than twice that of current IceCube optical modules.

The current baseline PMT for the IceCube Upgrade mDOM is a 3 inch PMT from Hamamatsu. But a 3.5 inch PMT from HZC Photonics (HZC onward) is also under consideration for possible use in the mDOM. This work contains characterization results for both PMT models. Tested PMT quantities include the PMT gain, timing properties and quantum efficiency. Given the low temperatures down to -20\,$^{\circ}$C in the deep Antarctic ice where the PMTs will operate, special attention is given to dark noise rates as a function of temperature. 

Section \ref{sec:setup} describes the PMT models and the experimental setup in more detail. In section \ref{sec:pulsechar}, various room-temperature measurements of PMT characteristics are presented and compared for the two PMT models. Section\,\ref{sec:tempdependent} summarizes temperature-dependent measurements, focusing on dark noise rates in particular. Parts of these proceedings are a condensed version of and a comparison between the PMT characterization results presented for the Hamamatsu PMT in \cite{ICU_Hamamatsu} and the HZC PMT in \cite{ICU_HZC}.

\section{PMT Models}\label{sec:setup}

\begin{figure}[tb]
\centering
\hfill
\includegraphics[trim=1.3cm 22cm 11cm 1cm, clip, width=0.49\textwidth]{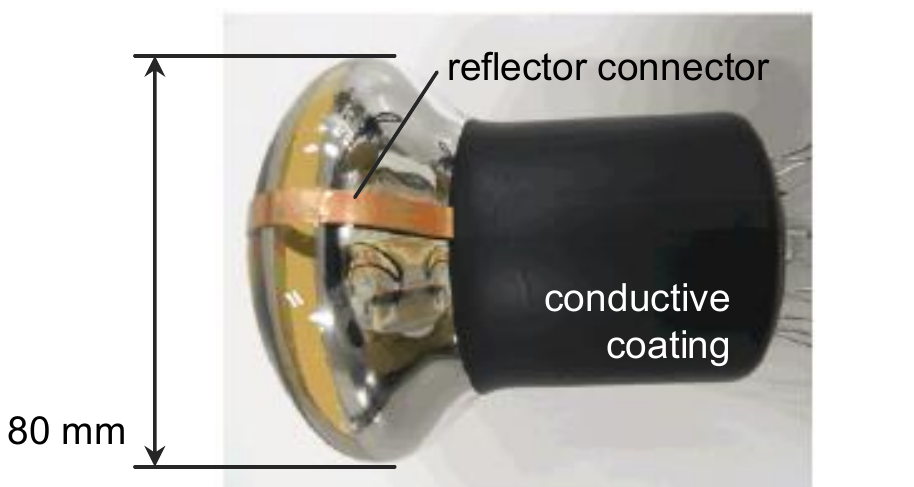}
\hfill
\includegraphics[trim=0 0 0 0, clip, width=0.33\textwidth]{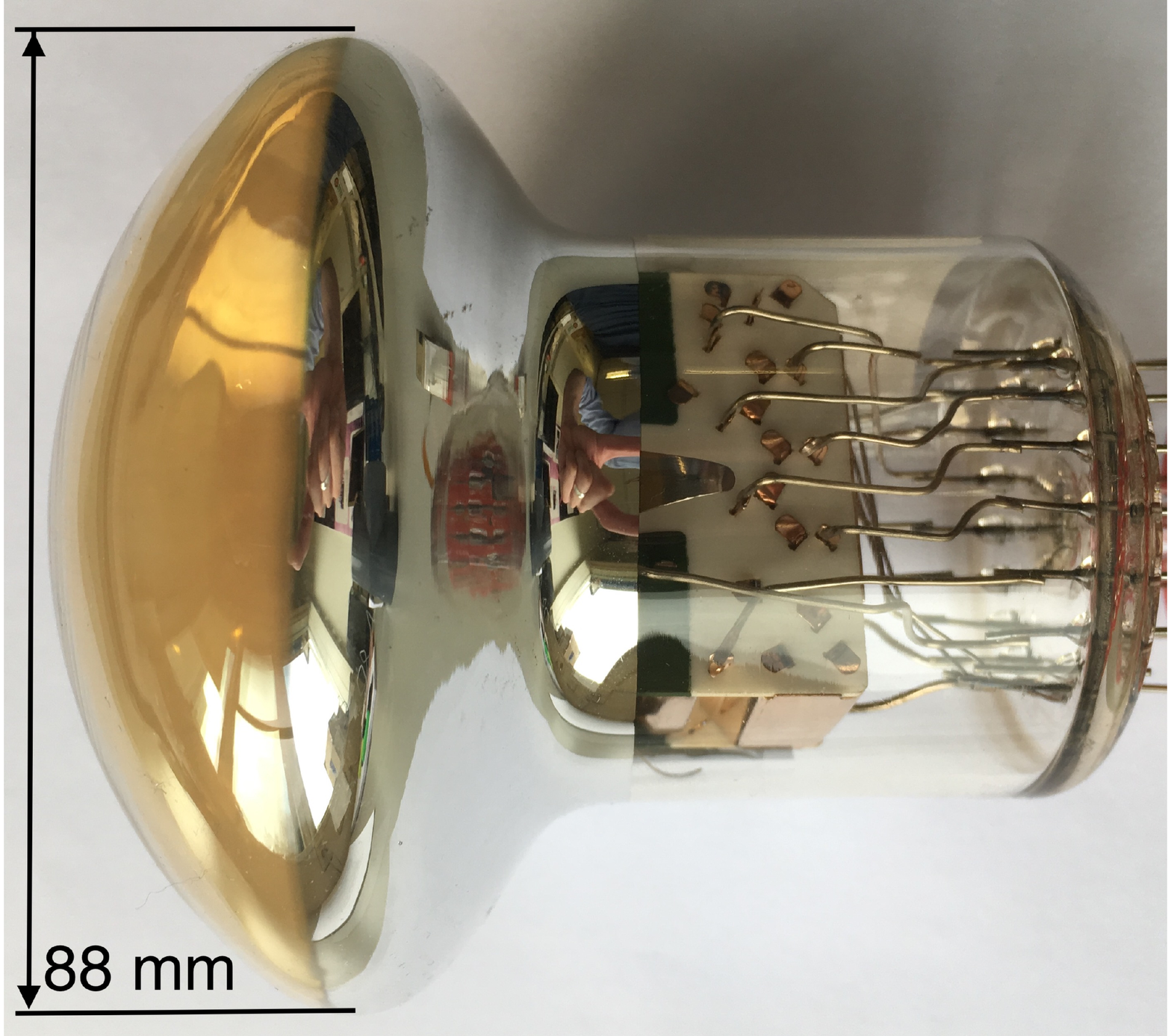}
\hfill
\hfill
\caption{Candidate PMTs for the IceCube Upgrade mDOM. Left: Hamamatsu R12199-01 MOD HA. Right: HZC XP82B2F.}
\label{fig:pmts}
\end{figure}

The two different PMT models (shown in figure~\ref{fig:pmts}) that have been characterized for possible use in the IceCube Upgrade mDOM are:
\begin{itemize}
\item Hamamatsu R12199-01 MOD HA\footnote{This model is based on the Hamamatsu R12199-02 that is currently used in the KM3NeT optical modules.} (100 pieces tested): 80 mm diameter, 10 dynode stages, borosilicate glass, bi-alkali photocathode, reduced tube length of 93 mm to fit mDOM spatial constraints, conductive HA coating to reduce dark noise for PMTs operated at negative high voltage (HV), copper strip connecting optical reflector rings to photocathode potential to reduce dark noise rates;
\item HZC XP82B2F (45 pieces tested): 88 mm diameter, 10 dynode stages, borosilicate low-K glass, bi-alkali photocathode, pin-compatible to and increased photosensitive area of 20-30\% compared to the R12199-01 MOD HA.
\end{itemize}

Both PMT models were operated using passive, negative polarity HV bases. The PMTs are operated in pulse mode and the PMT base signal output (capacitively coupled to the anode) is connected to an oscilloscope without amplification using a coaxial cable and 50\,$\Omega$ termination for visualization and analysis of the PMT pulse characteristics.

For all measured quantities except dark noise rates, an external trigger was used. The PMT was illuminated with a pulsed LED\footnote{PicoQuant PDL 800B} that was triggered by a pulse generator\footnote{Agilent 33220A} and also triggered the oscilloscope. The LED power was adjusted to obtain a mean illumination level of 0.1 photoelectrons (PE) per pulse. In order to achieve a homogeneous illumination of the photocathode, a diffuser was placed in front of the light source at sufficient distance from the PMT. 

\section{PMT Characteristics at Room Temperature}\label{sec:pulsechar}


\subsection{Charge Distribution and Gain Calibration}\label{sec:gain}
\begin{figure}[ht]
\includegraphics[trim = 0 0 0cm 0cm, clip, width=0.5\textwidth]{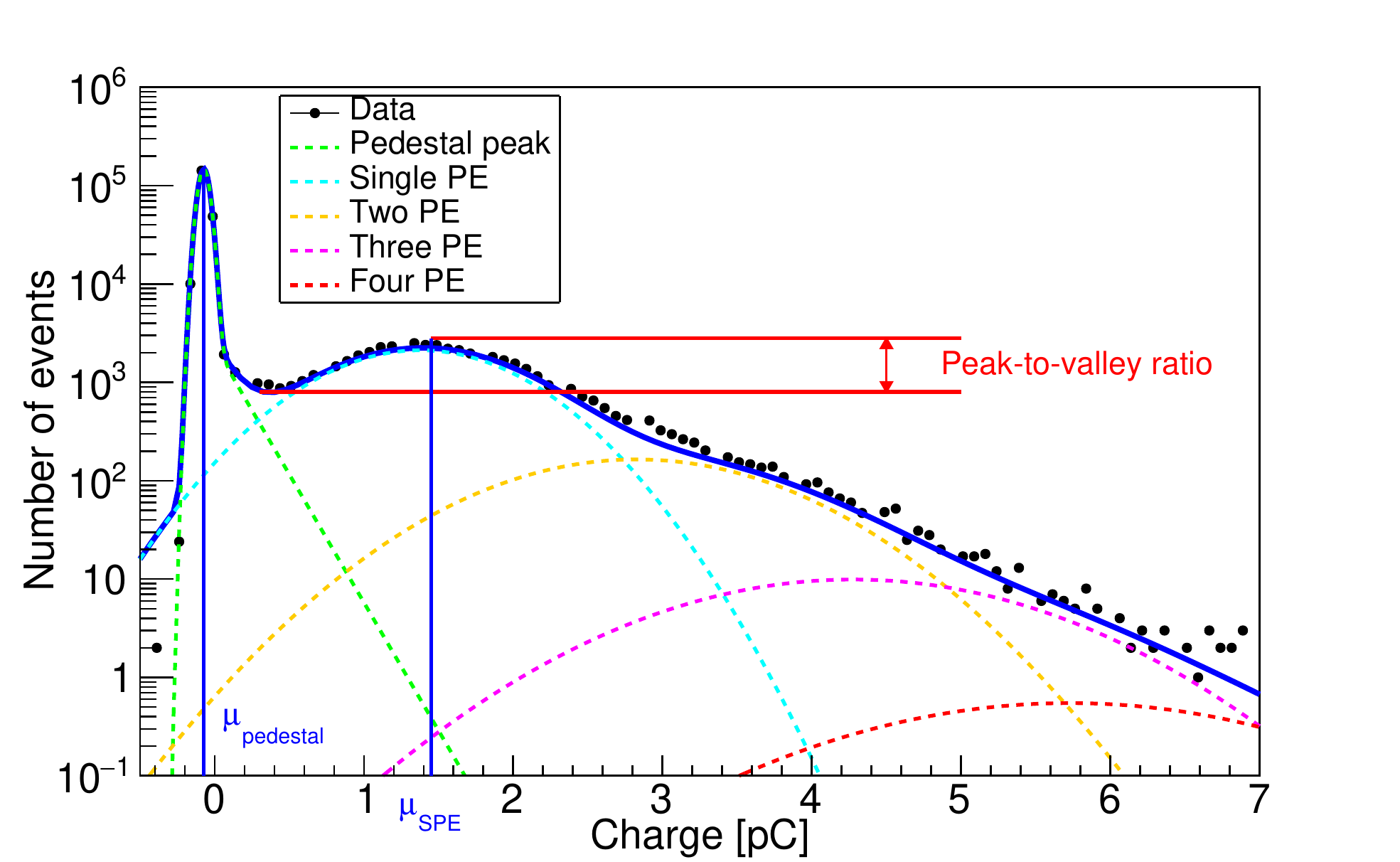}
\hfill
\includegraphics[trim=0 0 2cm 2cm,clip,width=0.47\textwidth]{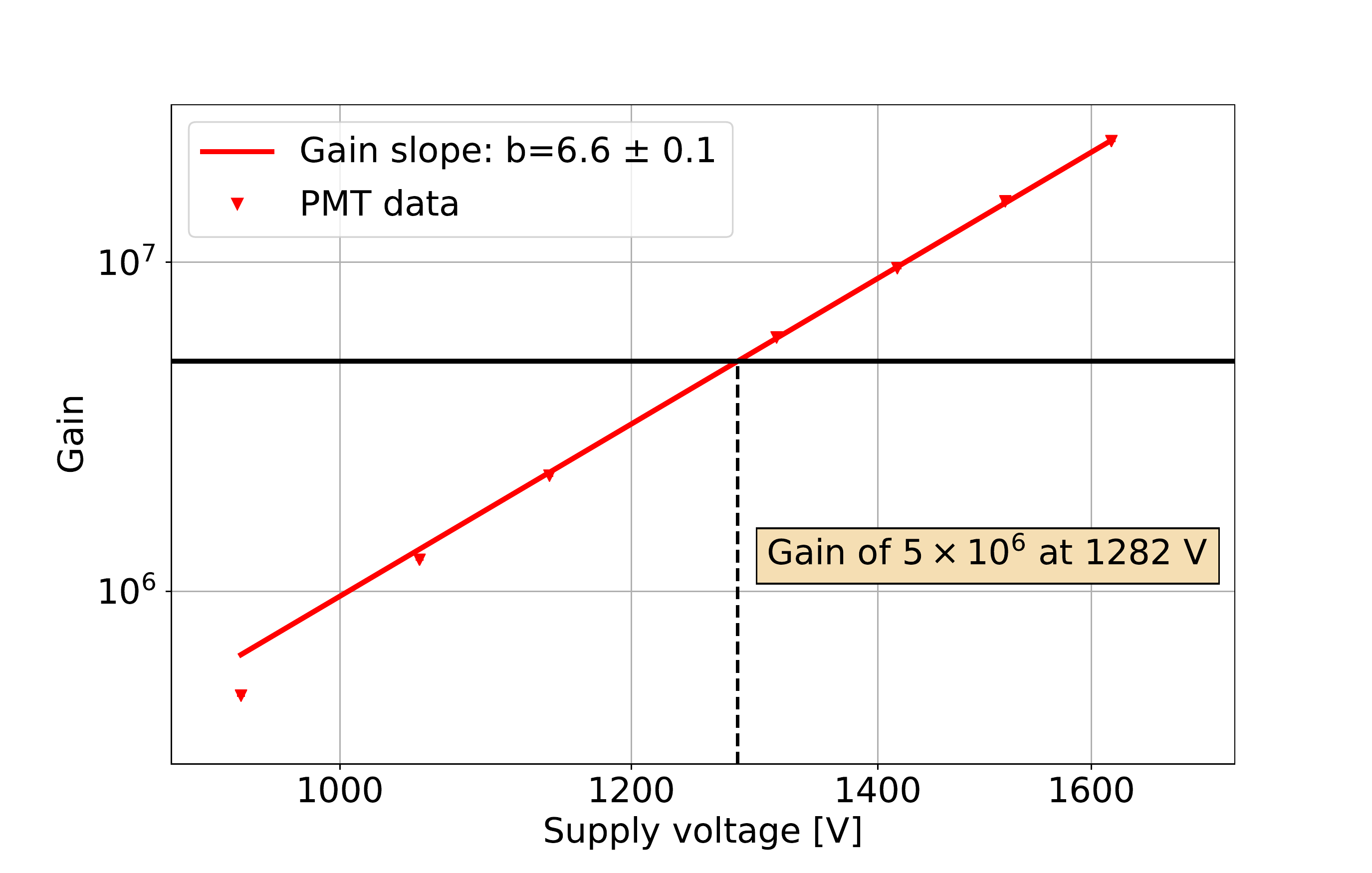}
\caption{Left: the charge distribution is a histogram of time-integrated PMT pulse shapes divided by the oscilloscope's input impedance, hence every entry in the histogram is equal to the collected charge in a pulse. The contributions from the pedestal peak, a single photoelectron (PE) and multiple photoelectrons are shown as dotted lines in various colours. The definition of the peak-to-valley ratio is indicated in red and the position of the fitted pedestal and SPE peak for calculation of the gain are shown in blue. Right: gain calibration curve for one particular HZC PMT.}
\label{fig:peandgain}
\end{figure}

The time integral of a PMT pulse as measured by the oscilloscope divided by the oscilloscope's input impedance (50\,$\Omega$) equals the charge collected by the PMT. By making a histogram of these time-integrated pulses, one obtains the charge distribution as shown in figure \ref{fig:peandgain} on the left. This distribution has various contributions, the largest being the pedestal peak corresponding to zero photoelectrons being present in the waveform. At higher collected charge, multiple (Gaussian) peaks contribute to the charge distribution, corresponding to one photoelectron in the PMT pulse, two photoelectrons, etc. The charge distribution is modelled according to \cite{PE_fit}.

\begin{figure}[ht]
\centering
\includegraphics[width=0.45\textwidth]{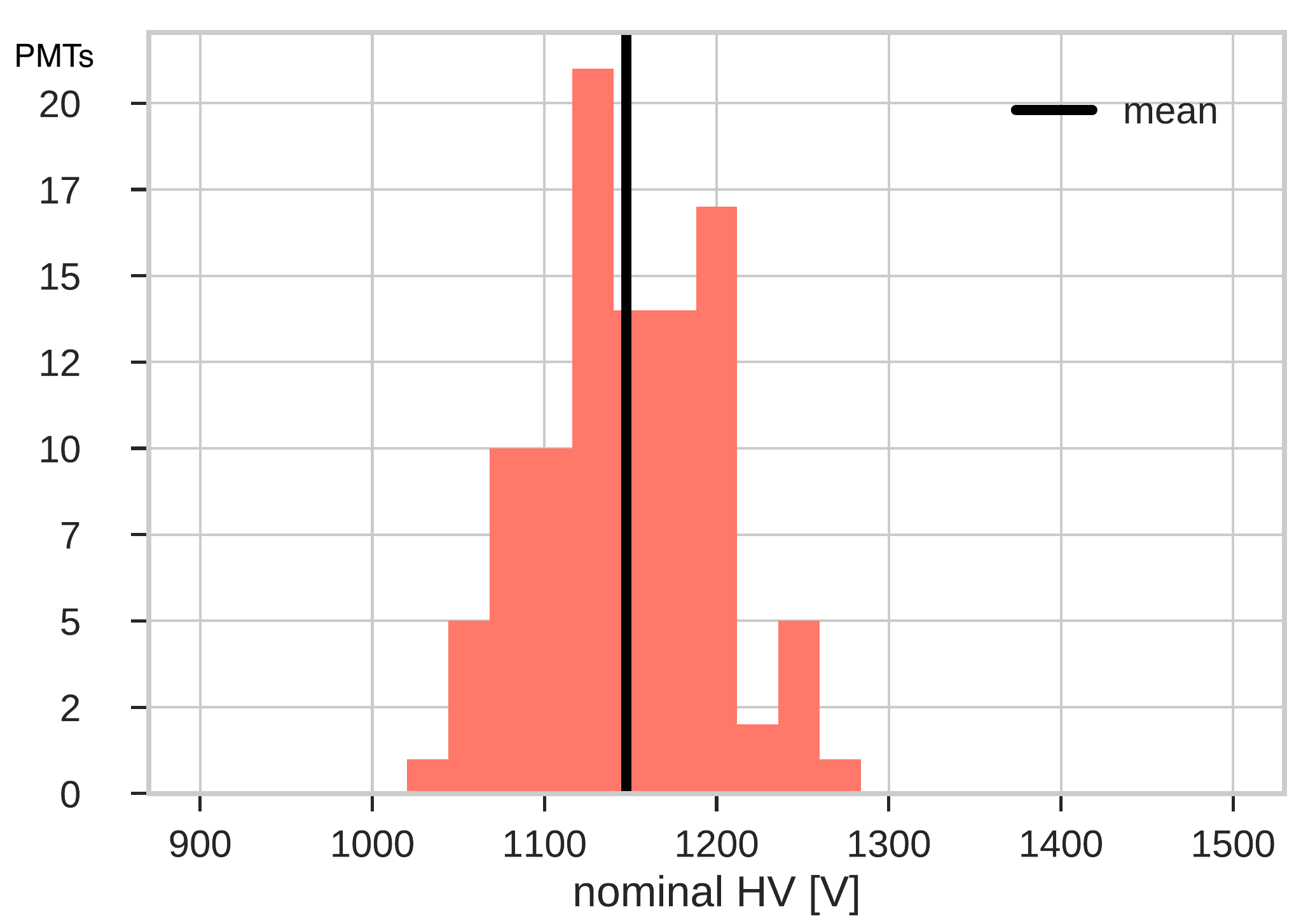}
\hfill
\includegraphics[width=0.45\textwidth]{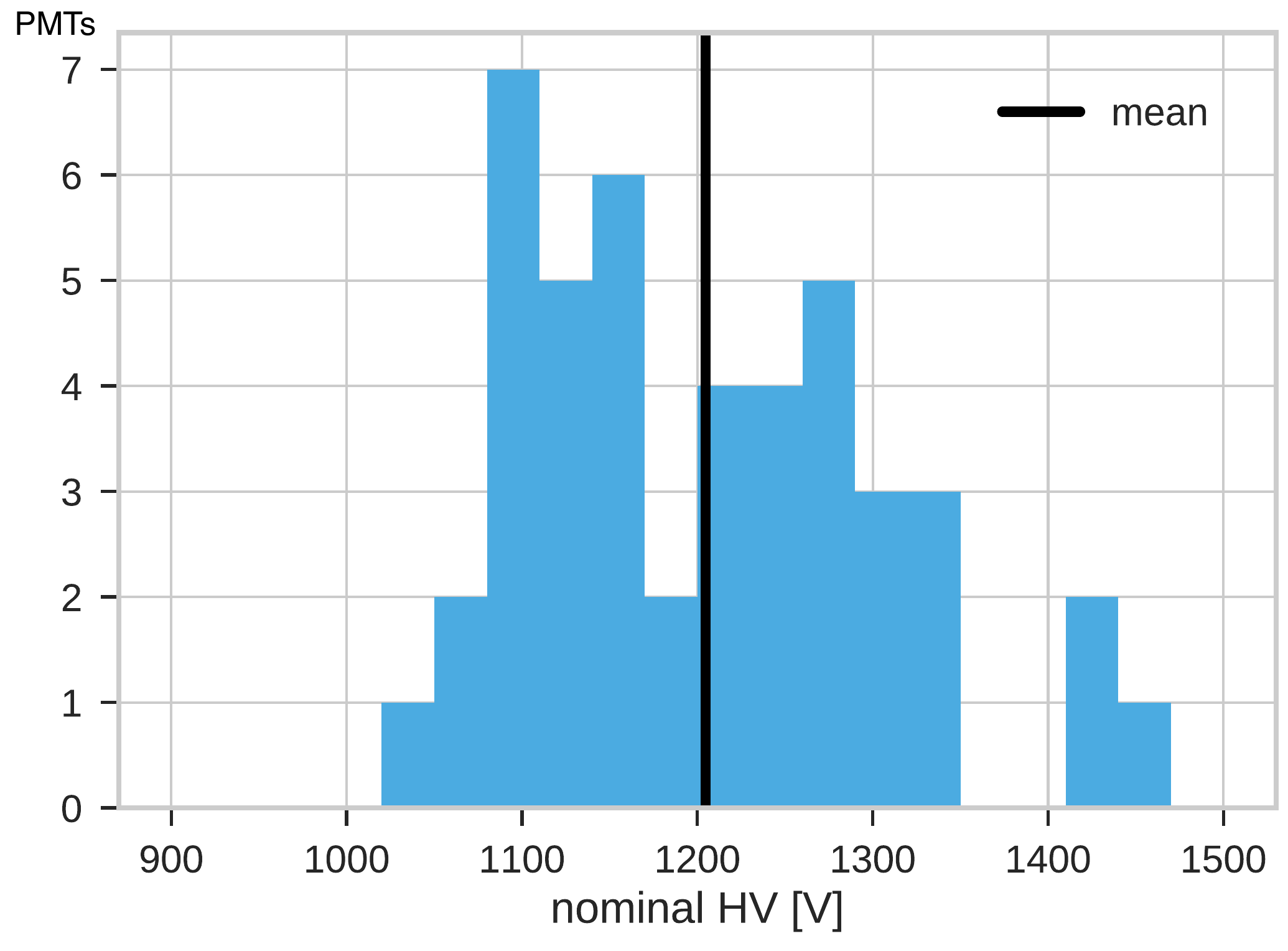}
\caption{Calibrated HV values for the 100 Hamamatsu (left) and 45 HZC (right) PMTs at a gain of $5 \times 10^6$.}
\label{fig:gainvshv}
\end{figure}

The PMT gain is defined as the average number of detected electrons at the anode as a result of the amplification in the dynode structure of a single photoelectron (SPE) that hits the first dynode. This number corresponds to the charge difference $\mu_{\mathrm{SPE}}-\mu_{\mathrm{pedestal}}$ between the position of the fitted SPE peak (corresponding to the maximum (mode) of the SPE peak in the charge distribution) and the position of the fitted pedestal peak divided by the electron charge $Q_{\mathrm{e}}$:
\begin{equation}
G = \frac{\mu_{\mathrm{SPE}}-\mu_{\mathrm{pedestal}}}{ Q_\mathrm{e}} \quad .
\end{equation}
By calculating the gain as a function of the supply voltage, the resulting gain calibration curve can be fitted with a power law $a\cdot x^{b}$. An example of a PMT gain calibration curve is shown in figure \ref{fig:peandgain} on the right. Figure \ref{fig:gainvshv} shows the gain-calibrated HV values for the 100 Hamamatsu and 45 HZC PMTs at a baseline nominal gain of $5 \times 10^6$. The mean of the histograms and the minimum and maximum values are summarized in Table \ref{tab:results_no_dark_rates}.

The peak-to-valley ratio is defined as the number of hits per bin at the maximum (mode) of the SPE peak divided by the number of hits per bin at the minimum between the pedestal peak and the SPE peak as shown in figure \ref{fig:peandgain} on the left. This quality parameter combines the performance of the PMT in terms of baseline electronic noise and SPE resolution. The comparison of peak-to-valley measurements between the two PMT models are summarized in table \ref{tab:results_no_dark_rates}.

\subsection{Timing Characteristics}
\begin{figure}[ht]
	\centering
	\includegraphics[width=0.5\textwidth]{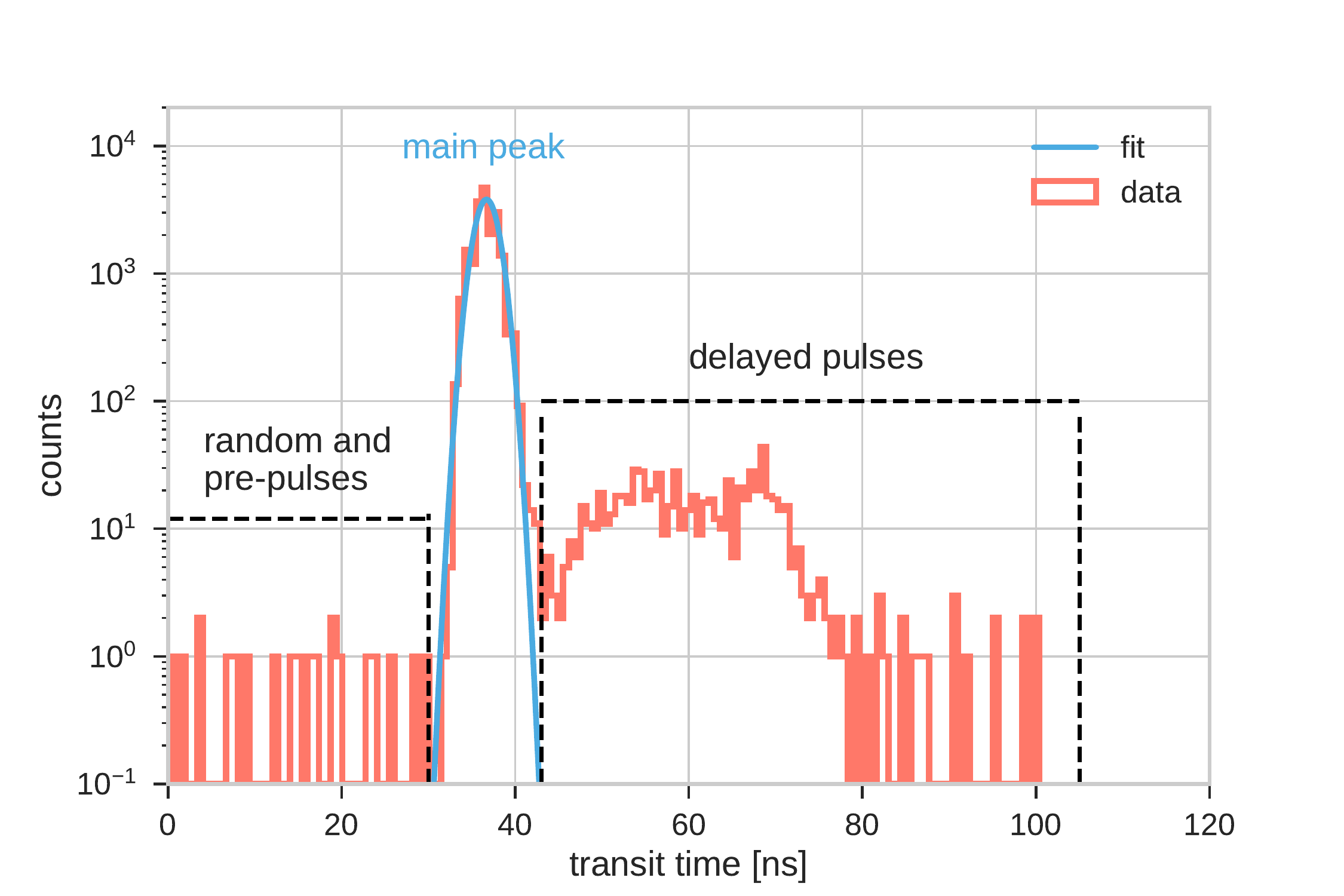}
	\hfill
	\includegraphics[width=0.49\textwidth]{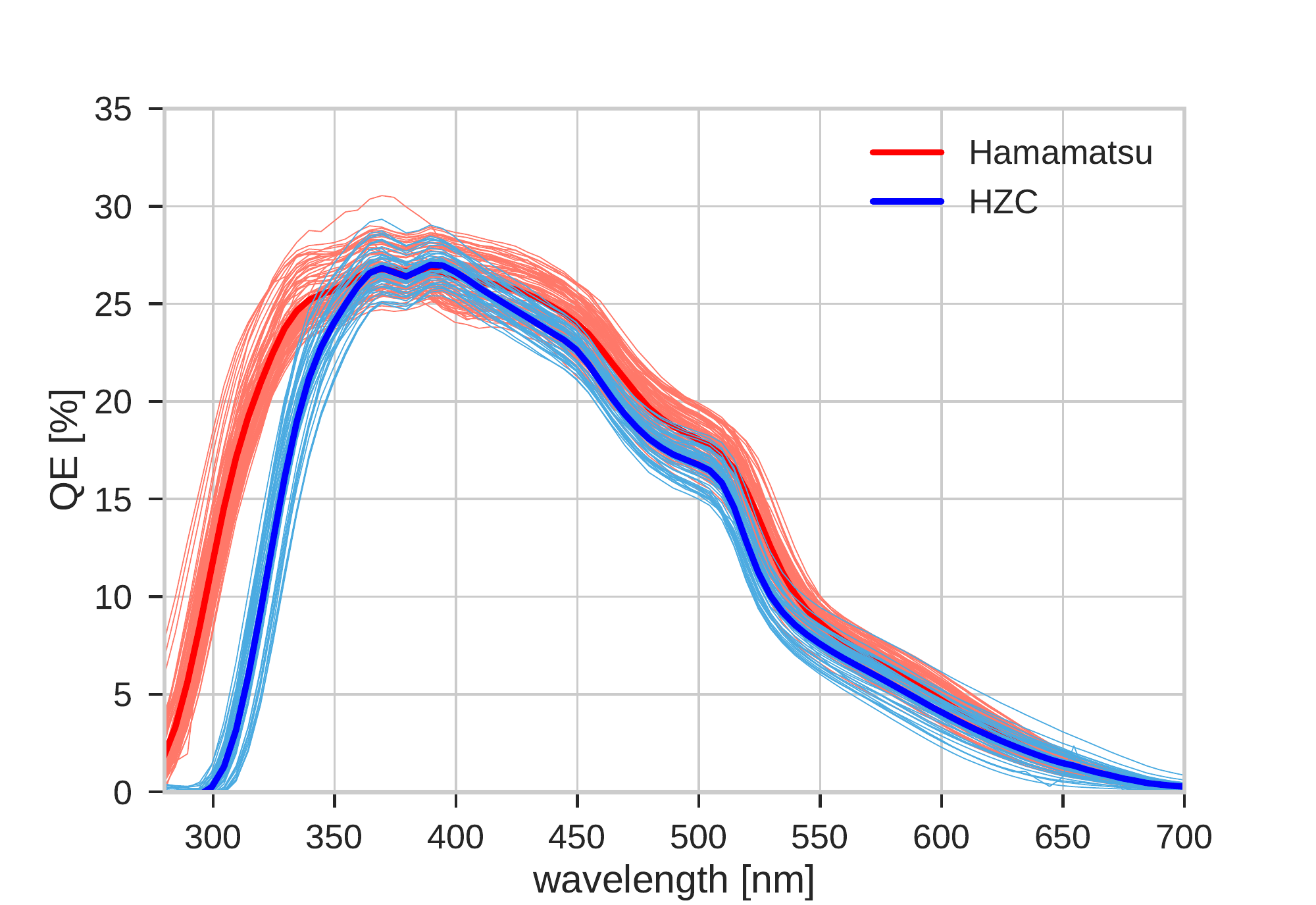}
	\caption{Left: transit time histogram for one particular PMT, showing a Gaussian fit to the main peak. The fitted standard deviation is defined as the transit time spread. Also shown are the early pulses and delayed pulses before and after the main peak. Right: QE measurements for the two PMT models.}
\label{fig:pre_delayed_tts_and_qe}
\end{figure}

The PMT transit time is the time between photoelectron emission (for which the pulse time of the external pulser can be used as a proxy) and the moment when the PMT signal reaches its maximum amplitude. Using a trigger level threshold of 0.25 PE, figure \ref{fig:pre_delayed_tts_and_qe} shows a transit time histogram for one particular PMT. The single-photon time resolution of a PMT is quantified by the transit-time spread (TTS), defined as the standard deviation of a Gaussian distribution fitted to the main peak in the transit time histogram. 

As seen in figure \ref{fig:pre_delayed_tts_and_qe}, there are additional contributions to the transit time histogram before and after the main peak. These are called early pulses (some of which might be genuine pre-pulses, others are random pulses) and delayed pulses respectively\footnote{For a description of early pulses and delayed pulses and their physical mechanisms, see \cite{Early_Delayed_Pulses}}. The early and delayed pulse numbers are calculated as a fraction of the total number of hits in a time window of 10 to 40 ns before and 15 to 100 ns after the hits in the main peak, respectively. 

The TTS, early pulse fraction and delayed pulse fractions for the two PMT models are summarized in table \ref{tab:results_no_dark_rates}. 

\subsection{Quantum Efficiency}\label{sec:QE}
The PMT quantum efficiency (QE) was measured by guiding monochromatic light selected via a monochromator from the spectrum of a continuous light source to the PMT photocathode, illuminating the innermost region of approximately 2 cm. The resulting photocathode current was collected via a dedicated base that shorts the electron multiplier system and calibrated with the current from a reference photo-diode exposed to the same photon flux. The resulting QE spectra are presented in figure \ref{fig:pre_delayed_tts_and_qe} and QE values at two benchmark points are summarized in table \ref{tab:results_no_dark_rates}.

\section{Temperature-Dependent PMT Characteristics}\label{sec:tempdependent}
Low-temperature behaviour of PMTs is of particular importance for possible use in future IceCube modules because of the ambient temperatures down to -20\,$^{\circ}$C in the deep Antarctic ice where the PMTs must ultimately operate. Temperature-dependent PMT characterization studies are performed by placing a subset of PMTs in a temperature-controlled dark freezer. Twelve Hamamatsu PMTs and three HZC PMTs were tested at low temperatures.

\begin{figure}[ht]
	\centering
	\hfill
	\includegraphics[width=0.49\textwidth]{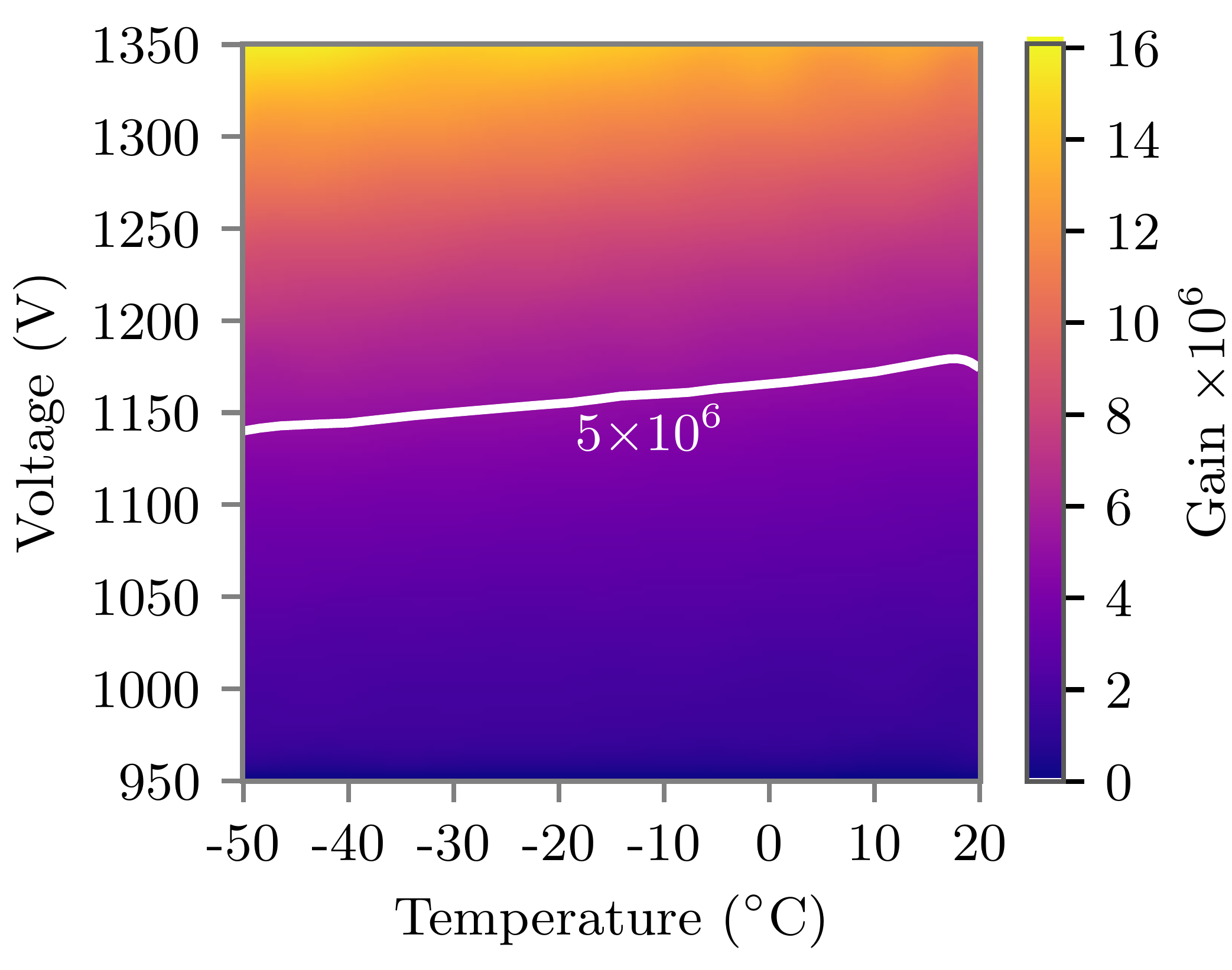}
	\hfill
	\includegraphics[width=0.49\textwidth]{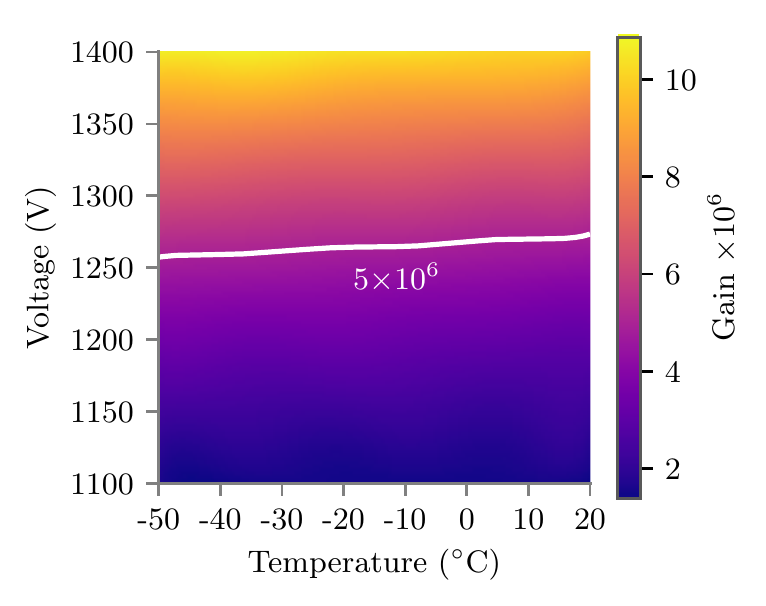}
	\hfill
	\hfill
	\caption{Gain and operating voltage as a function of temperature for one particular Hamamatsu PMT (left, plot taken from \cite{ICU_Hamamatsu}) and a HZC PMT (right).}
\label{fig:T_gain}
\end{figure}

The low-temperature tests on the two Hamamatsu PMTs showed no dependence of the PMT pulse shape and TTS on temperature. There is an effect on gain though (and therefore on operating voltage), as can be seen in figure \ref{fig:T_gain}. To compensate, the operating voltages of the PMTs were adjusted as a function of temperature in order to keep the PMT gain constant.

\subsection{Temperature Dependence of Dark Noise Rates}
While measuring dark noise rates, no external light source is used and PMTs are kept in the dark with HV on for several hours before the start of the measurements to counteract the inadvertent increase of dark noise due to light exposure while handling the PMTs. A trigger threshold of 0.25 PE was used as well as an artificial dead time of 0.5 $\mu$s.

\begin{figure}[ht]
	\centering
	\hfill
	\includegraphics[width=0.49\textwidth]{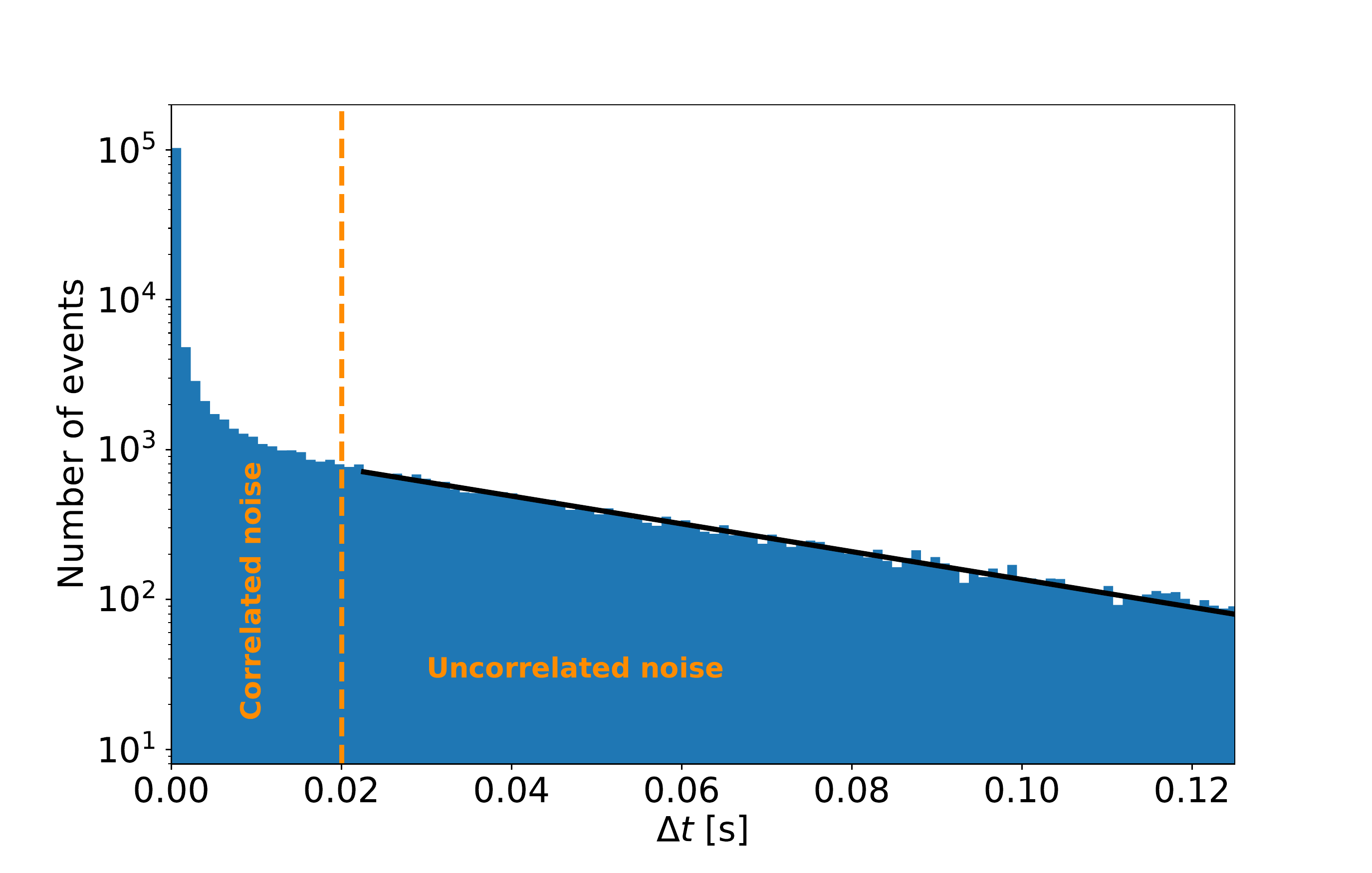}
	\hfill
	\includegraphics[trim = 0 0 0 2cm, clip, width=0.5\textwidth]{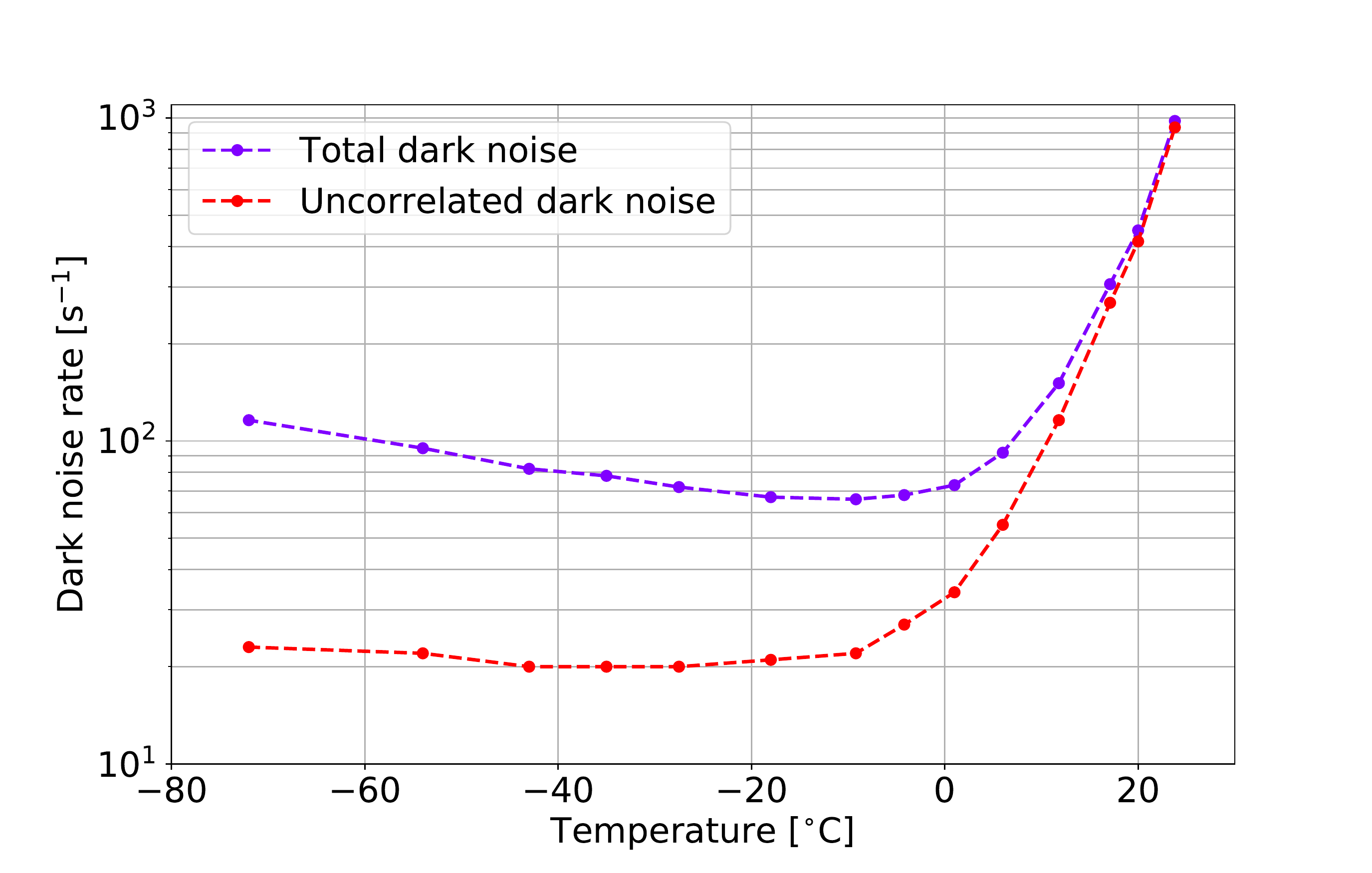}
	\hfill
	\hfill
	\caption{Left: distribution of hit time differences for one particular HZC PMT at roughly -20$^{\circ}$C: correlated dark noise hits occur at small hit time differences and uncorrelated dark noise hits occur at larger hit time differences. Right: Total dark noise rates and uncorrelated dark noise rates as a function of temperature for the same PMT.}
\label{fig:T_noise}
\end{figure}

The total dark noise rate is defined as the total number of pulses divided by the run time. By binning the time difference between subsequent PMT pulses, various contributions to the total dark noise can be distinguished. At large hit time differences, there is an exponentially falling contribution called the uncorrelated noise. However, at small time differences there is an additional (non-exponential) contribution to the total dark noise rate as shown in figure \ref{fig:T_noise} on the left. This contribution is called the correlated noise and is believed to originate from radioactive decays and scintillation in the PMT glass \cite{Lew}.

The total dark noise and uncorrelated dark noise as a function of temperature for one particular HZC PMT is shown in figure \ref{fig:T_noise} on the right. At room temperature, the dark noise is dominated by electrons released in the photocathode due to thermionic emission. As the temperature drops, the thermionic component of the dark noise decreases until the total dark noise rate is well below 100 $s^{-1}$ at -20$^{\circ}$C. As the temperature drops even further, there is a slight increase of the dark noise rates, which is probably caused by the rise of the scintillation yield of the glass.

Dark noise rate measurements for both PMT models at room temperature and a benchmark temperature of -20$^{\circ}$C are summarized in table \ref{tab:dark_rates}. Both PMT models have excellent noise behaviour with dark noise rates well under 100 $s^{-1}$ at the low-temperature benchmark. The larger dark noise rates of the HZC PMTs can be explained by their larger photocathode area and the fact that they were operated at a higher gain of $1 \times 10^7$ in these particular measurements. In addition, one of the three tested HZC PMTs had particularly high room temperature dark noise rates, strongly biasing the (low statistics) averages in table \ref{tab:dark_rates}.

\section{Conclusions}
Two PMT models were characterized for possible use in the future IceCube Upgrade mDOM: the 3 inch R12199-01 HA MOD PMT from Hamamatsu and the 3.5 inch XP82B2F PMT from HZC Photonics. Various PMT characteristics were measured at room temperature and a comparison between the two PMT models is made in table \ref{tab:results_no_dark_rates}. 

\begin{table}[ht]
\begin{centering}
\begin{tabular}{|lcc|}
\hline
& \textbf{Hamamatsu} & \textbf{HZC} \\
& R12199-01 HA MOD & XP82B2F \\
& 100 pieces tested & 45 pieces tested \\
\hline
HV at gain of $5 \times 10^6$ [V] & 1148 (1038 - 1270) & 1205 (1048 - 1451) \\
Peak-to-valley ratio & 3.52 (2.65 - 4.85) & 3.28 (2.56 - 4.38) \\
Transit time spread [ns] & 1.49 (1.35 - 1.66) & 1.91 (1.31 - 2.34) \\
Early pulses [\%] & 0.02 (0.00 - 0.13) & 0.26 (0.00 - 0.68) \\
Delayed pulses [\%] & 2.19 (1.42 -3.28) & 3.58 (2.28 - 4.72) \\
QE at 390 nm [\%] & 26.8 (24.8 - 29.1) & 27.0 (25.4 - 29.0) \\
QE at 470 nm [\%] & 21.1 (18.6 - 23.5) & 19.3 (17.8 - 20.7) \\
\hline
\end{tabular}
\caption{Mean histogram values for various PMT characterization measurements at room temperature. The numbers in parentheses are the minimum and maximum values of the histograms.}
\label{tab:results_no_dark_rates}
\end{centering}
\end{table}

Since the IceCube Upgrade mDOMs will operate in the deep ice of Antarctica, temperature-dependent dark noise rate measurements were performed as well. The results for the two PMT models are compared in table \ref{tab:dark_rates}.

\begin{table}[ht]
\begin{centering}
\begin{tabular}{|lcc|}
\hline
& \textbf{Hamamatsu} & \textbf{HZC} \\
& R12199-01 HA MOD & XP82B2F \\
& 12 pieces tested & 3 pieces tested \\
\hline
Total DNR [$\mathrm{{s}}^{{-1}}$] at 20$^{\circ}$C & 191 & 804 \\
Uncorrelated DNR [$\mathrm{{s}}^{{-1}}$] at 20$^{\circ}$C & 120 & 762 \\
Total DNR [$\mathrm{{s}}^{{-1}}$] at -20$^{\circ}$C & 35 & 67 \\
Uncorrelated DNR [$\mathrm{{s}}^{{-1}}$] at -20$^{\circ}$C & 12 & 22 \\
\hline
\end{tabular}
\caption{Average total dark noise rate and uncorrelated dark noise rate at room temperature (first two rows) and at -20$^{\circ}$C (last two rows). The HZC PMTs were operated at a gain of $1\times10^7$ (compared to $5\times10^6$ for the Hamamatsu PMTs), which contributes to the observed dark noise rate differences.}
\label{tab:dark_rates}
\end{centering}
\end{table}

In conclusion, the results for both PMT models are in good agreement with the specifications as provided by the manufacturers and show excellent noise behaviour at the low-temperatures relevant for possible use in future IceCube optical modules. More detailed information on the PMT characterization studies can be found in \cite{ICU_Hamamatsu} for the Hamamatsu PMTs and in \cite{ICU_HZC} for the HZC Photonics PMTs.

\bibliographystyle{ICRC}
\bibliography{references}

\providecommand{\href}[2]{#2}\begingroup\raggedright\begin{thebibliography}{1}

\bibitem{IceCube}
{\bf IceCube} Collaboration, M.~G. Aartsen et~al., {\em JINST} {\bf 12} (2017)
  P03012.

\bibitem{IceCubeUpgrade}
{\bf IceCube Gen2} Collaboration, M.~G. Aartsen et~al.,
  \href{http://arxiv.org/abs/1412.5106}{{\tt arXiv:1412.5106}}.

\bibitem{KM3NeT}
{\bf KM3NeT} Collaboration, S.~Adri\'{a}n-Mart\'{i}nez et~al., {\em J.\,Phys.}
  {\bf G43} (2016) 084001.

\bibitem{KM3NeTDOM}
R.~Bruijn and D.~van Eijk,  \pos{PoS(ICRC2015)1157} (2016).

\bibitem{ICU_Hamamatsu}
{M. A. Unland Elorrieta} et~al., {\em Journal of Instrumentation} {\bf 14}
  (2019) P03015.

\bibitem{ICU_HZC}
D.~van Eijk et~al., {\em Journal of Instrumentation} {\bf 14} (2019) P07009.

\bibitem{PE_fit}
E.~Bellamy et~al., {\em Nucl. Instrum. Methods Phys. Res. A} {\bf 339} (1994)
  468 -- 476.

\bibitem{Early_Delayed_Pulses}
B.~Lubsandorzhiev et~al., {\em Nucl. Instrum. Methods Phys. Res. A} {\bf 442}
  (2000) 452 -- 458.

\bibitem{Lew}
L.~Classen, {\em PhD thesis Friedrich-Alexander-Universit{\"a}t
  Erlangen-N{\"u}rnberg (FAU)} (2001).

\end{thebibliography}\endgroup

\end{document}